\documentclass[aps,pra,twocolumn,superscriptaddress,nofootinbib]{revtex4-2}

\usepackage{graphicx}
\usepackage{amsmath, amssymb, bm, braket, amsfonts}
\usepackage{xcolor}
\usepackage{hyperref}
\hypersetup{
    colorlinks=true,
    linkcolor=blue
}

\newcommand{\freiburg}{Physikalisches Institut, Albert-Ludwigs-Universit\"at Freiburg,\\ Hermann-Herder-Stra\ss e 3, D-79104 Freiburg, Germany}
\newcommand{\eucor}{EUCOR Centre for Quantum Science and Quantum Computing, Albert-Ludwigs-Universit\"at Freiburg,\\ Hermann-Herder-Stra\ss e 3, D-79104 Freiburg,  Germany}

\newcommand{\nict}{Advanced ICT Research Institute, National Institute of Information and Communications Technology,\\ 4-2-1 Nukui-Kitamachi, Koganei, Tokyo 184-8795, Japan}

\newcommand{\tus}{Research Institute for Science and Technology, Tokyo University of Science,\\ 1-3 Kagurazaka, Shinjuku-ku, Tokyo 162-8601, Japan}

\begin{document}

\title{Is the most random pattern random? Maximizing localization in a two-dimensional lattice with engineered disorder}
	\author{Morgan Berkane}\affiliation{\freiburg}\affiliation{\eucor}
    \author{Sahel Ashhab}\affiliation{\nict}\affiliation{\tus}
    
\date{\today}

\begin{abstract}
We investigate localization in two models: a single particle in a two-dimensional square lattice described by the tight binding  Hamiltonian, and a two-dimensional square qubit lattice. It is well-known that Anderson localization 
occurs under suitable conditions in which the system parameters are chosen randomly from some statistical distribution. We propose a situation in which the parameters, specifically the on-site energies, are carefully chosen in such a way that a localization-quantifying parameter is maximized. We demonstrate the optimization procedure with numerical calculations in which the engineered localization significantly exceeds the average localization caused by a random distribution of the on-site energies. We explore the relation between spatial patterns and localization efficiency. Furthermore, we use perturbation theory to gain insight into the localization mechanism and obtain an improved cost function for optimization calculations, leading to enhanced localization in both the single-particle and full Hilbert spaces. Although large-scale simulations for qubit lattices are computationally infeasible, we use small-system simulations to demonstrate that results obtained using the single-particle tight binding  model can be adapted to identify optimal settings for qubit lattice systems to achieve maximum decoupling between the qubits, which can be valuable for optimizing the idle-state settings on a quantum processor.
\end{abstract}

\maketitle

\section*{Introduction}

Localization phenomena play a crucial role in understanding and utilizing transport and dynamics in physical systems\cite{Anderson1958,Abrahams1979Scaling,anderson201050,Basko2006MBL,Abrahams1979Scaling,Genack1991}.Of particular interest to us in this work is Anderson localization, which arises as a result of disorder in the potential energy landscape of a quantum system, with the model of a quantum particle hopping between lattice sites being the standard paradigm for studying this phenomenon\cite{Ashhab2018EffectiveHopping,Billy2008Anderson,AshcroftMermin1976,Madelung1978,JakschZoller2005,Gurarie2009}. A large number of studies have investigated various relevant questions, including the localization-delocalization phase transition \cite{Berke2022TransmonChaos,Basko2006MBL,madronero2006quantum}, the role of system dimensionality in this transition and properties of the localized states\cite{evers2000fluctuations}.

The conventional approach to thinking about disorder-induced localization is to assume that the disorder is caused by uncontrolled interactions and experimental imperfections that create a random potential energy landscape, which scrambles the phase of a propagating wave, causing reflections and preventing the particle from propagating away from its initial location.

With the rapid advances in quantum technologies, we are having ever increasing control over the parameters of quantum systems. Impressive experiments have demonstrated the ability to control the parameters of every site in a lattice, including atoms in optical lattices and superconducting Josephson-junction-based lattices\cite{Bakr2009,MegrantChen2025}

These recent technological advances raise the possibility to induce Anderson localization in a controlled way. In other words, one can design a potential that mimics a random potential but is designed to achieve a specific purpose, e.g.~a specific level of localization\cite{SanchezPalencia2010}. In particular, we are interested in maximizing the localization, which is a desirable situation in a variety of physical settings\cite{izrailev1999localization,deMoura1998,Dunlap1990}. It is then possible to perform optimization calculations to identify the potential that achieves the desired goal\cite{SanchezPalencia2010,izrailev1999localization}. Interestingly, in this case, the problem could be formulated as searching for the most random potential, even though this most random potential is not random at all, but rather specified by the requirement of maximizing the localization. It is interesting to note here that one can also perform the opposite task of enabling transport in the presence of disorder using optimized control pulses \cite{ashhab2015}.

In recent years, a topic related to localization has emerged in the field of quantum computing\cite{heyl2019quantum}: In a variety of physical architectures used to build quantum processors, e.g.~in superconducting devices, interactions between neighboring qubits cannot be turned off completely. Considering that the interaction between any given pair of neighboring qubits needs to be activated for only a short period of time during the implementation of a quantum protocol, one must find a way to effectively decouple qubits during the idle periods when they are not being actively used. The standard approach to achieve effective suppression of always-on interactions is to strongly detune neighboring qubits from each other. While the physics of this effective decoupling is straightforward for two qubits, it becomes difficult to identify the optimal settings for effective decoupling when dealing with a qubit lattice. One of the main difficulties is the fact that the Hilbert space size grows exponentially with the number of qubits, rendering a simulation of the full quantum dynamics of the quantum processor practically impossible. It is therefore necessary to find alternative ways to identify optimal settings to maximize the effective decoupling between qubits during the idle periods \cite{Klimov2018Patent,Klimov2020Snake,Klimov2024Gates}. We tackle this problem by first noting that the decoupling between qubits is analogous to localization in particle transport and show that this analogy can be leveraged to infer favourable conditions for decoupling in a quantum processor from the much simpler task of solving a single-particle localization problem. Solving the single-particle transport problem scales only linearly in the number of sites, while solving the problem of decoupling an array of qubits involves a Hilbert space that scales exponentially in the number of qubits. We use $3 \times 3$ lattices to demonstrate the soundness of this approach, i.e.~the fact that a potential that maximizes localization in the signle-particle problem also leads to maximal decoupling in the much larger Hilbert space of a qubit lattice. It is important to keep in mind that the full dynamics of a qubit lattice exhibits phenomena that cannot be captured at the single-particle level, as was illustrated in a recent experiment using a superconducting processor \cite{lunkin2026}.

The remainder of this paper is organized as follows:  We start by introducing the tight binding  model\cite{fisher1989boson} and the qubit lattice model\cite{Kjaergaard2020Review} as well as the correlation functions that quantify the degree of localization in these models. Then we present the results of our numerical calculations that compare the degree of localization under various conditions and identify the conditions that lead to maximum localization. These results also demonstrate the similarity in the localization behaviour for the tight binding  and qubit lattice models, in spite of the vast difference in their complexity scaling. Furthermore, we present a perturbation theory apprach that provides an even simpler method to identify optimal conditions for localization. We then conclude with some final remarks.

\subsection*{Model and method}

\subsubsection{tight binding  Hamiltonian}

We consider the tight binding  Hamiltonian that describes particles hopping between the sites of a periodic lattice:
\begin{equation}
    \hat{H} = \sum_{i=1}^n W_i \hat{a}_i^\dagger \hat{a}_i + \sum_{\langle i,j \rangle} J_{ij} (\hat{a}_j^\dagger \hat{a}_i + \text{h.c.}),
    \label{eq:BHhamiltonian}
\end{equation}
where $\hat{a}_i$ annihilates a particle at site $i$, $\hat{a}_i^\dagger$ is the hermitian conjugate of $\hat{a}_i$ and hence creates a particle at site $i$, $W_i$ is the energy of site $i$, $J_{ij}$ (which we take to be real) is the hopping amplitude between sites $i$ and $j$, and $n$ is the number of sites in the lattice. The notation $\langle i,j \rangle$ indicates that we take the sum over nearest-neighbor pairs of sites.
The time evolution of the quantum state $\psi(t)$ is described by the unitary evolution $\ket{\psi(t)} = e^{-i \hat{H} t} \ket{\psi(0)}$. 

\subsubsection{Maximization of localization in the single-particle sector}

One type of localization that arises under the tight binding  Hamiltonian of Eq.~(\ref{eq:BHhamiltonian}) is the single-particle Anderson localization of a single particle\cite{nandkishore2015many}. The single-particle Hilbert space has dimension $n$ with basis $\mathcal{B}_{1p} = \{\ket{1,0,\dots,0}, \ket{0,1,0,\dots,0}, \dots, \ket{0,\dots,0,1}\}$.

To quantify single-particle localization, we introduce the site-averaged autocorrelation function
\begin{equation}
    \mathcal{C}(t) = \frac{1}{n} \sum_{i=1}^{n} 
    \bigl|\braket{\psi_i(0)|\psi_i(t)}\bigr|^2,
    \label{eq:autocorrelationfunction}
\end{equation}
where $\ket{\psi_i(0)} = \ket{0,\dots,1_i,\dots,0}$ denotes the initial state
with a single excitation localized at site $i$. Localization occurs when
$\mathcal{C}(t) \approx 1$ for all times, i.e.\ when the return
probability, averaged over all initial sites, remains close to unity and the
excitation essentially does not leave its original location. We note here that this is a strong definition of localization. A weaker definition would allow for the particle to move to neighboring sites but not to far-away sites\cite{evers2008anderson}. The strong definition is more relevant for the practical applications that we will consider below.

Our objective is to find the set of on-site energies $\{W_i\}$ that maximizes this localization parameter. We fixe $J_{ij}=J=1$ and we choose the on-site energies randomly from a uniform distribution $W_i \in [0,100]$, which is a realistic scenario for our purposes. To this end, we first define the cost function
\begin{equation}
    \mathcal{L}(\{W_i\}) = \sum_{k} \bigl[\mathcal{C}(t_k) - 1\bigr]^2,
    \label{eq:costfunction}
\end{equation}
In the numerical calculations, we take the $t_k$ values $[0,0.1,0.2,...,10]$. Note that, considering Eq.~(\ref{eq:autocorrelationfunction}), the cost function $\mathcal{L}(\{W_i\})$ includes averages over space and time.
We minimize $\mathcal{L}(\{W_i\})$ with respect to $\{W_i\}$ in order to maximize localization in the single-particle sector.

\subsubsection{Maximization of localization in a qubit lattice}

The second type of localization that we analyze in this work is that of an arbitrary-particle-number quantum states in a qubit lattice, which is relevant to quantum processors. In this case, we consider the Hilbert space in which each site can contain zero or one particle, representing the qubit states $\ket{0}$ and $\ket{1}$. An intuitive picture of this localization problem can be obtained by considering that the qubit lattice is in a separable state described by a product of single-qubit states. We then require that this product state remains intact and does not evolve into an entangled state. If we achieve this situation, we have effectively localized the quantum states and suppressed the effects of so-called residual interactions between the qubits. When dealing with two transversally interacting qubits that are detuned from each other, it is well known that the transverse interactions do not cause entangling dynamics to lowest order but cause a renormalization of the single-qubit energies, i.e.\ the on-site energies. We must therefore allow for such a renormalization in on-site energies. 
Different studies \cite{Klimov2020Snake,inoue2026,Zhang2025,Morvan2022,McKinney2024} proposed optimization strategies for determining suitable qubit configurations, typically relying on frequency-allocation procedures and local constraints between neighboring qubits across the lattice. In contrast, we employ a different approach tailored to our system, based on a global optimization of a localization-driven cost function.
Hence, we define the cost function
\begin{widetext}
\begin{equation}
\mathcal{L}_{FHS}= \sum_k \left(\left| \frac{{\rm Tr} \left[ \exp \left\{ -i t_k \hat{H} \right\} \exp \left\{ i \tau \hat{H}_{\rm eff} \left( \tilde{\omega}_1, \tilde{\omega}_2, \cdots, \tilde{\omega}_n \right) \right\} \right]}{2^n} \right|-1 \right)^2,
\label{eq:costfunction2}
\end{equation}
\end{widetext}
where
\begin{equation}
    \hat{H}_{\rm eff}\left( \tilde{\omega}_1, \tilde{\omega}_2, \cdots, \tilde{\omega}_n \right) = \sum_{i=1}^n \tilde{\omega}_i \hat{a}^\dagger_i \hat{a_i}.
\end{equation}
and $\tilde{\omega}_i$ are the renormalized on-site energies, which are initially unknown and need to be determined.
This cost function compares the exact evolution generated by $\hat H$ with
that of the diagonal, purely on-site Hamiltonian $\hat H_{\rm eff}$. Since
$\hat H_{\rm eff}$ is diagonal in the site basis, its dynamics only produces
local phase accumulation and does not move excitations between sites. The
normalized trace in Eq.~\eqref{eq:costfunction} therefore quantifies the overlap between the exact evolution under the full tight binding  Hamiltonian and such a purely local evolution. A small value
$\mathcal{L}_{FHS}\simeq 0$ indicates that $e^{-i t_k\hat H}$ is almost
diagonal in the Fock basis, i.e., the excitations remain localized and the
dynamics are well approximated by local on-site phase accumulation only. Higher values of $\mathcal{L}_{FHS}$, even after optimizing the renormalized on-site energies $\tilde{\omega}_i$, indicate increasing delocalization.

The minimization of the functionals is performed using a genetic algorithm, which efficiently explores the high-dimensional parameter space $\{w_1, \dots, w_N\}$. 
The optimization is repeated for multiple independent random initializations using the module \texttt{differential evolution} from \texttt{scipy.optimize} of Python, with the tolerance set to $10^{-6}$ and the maximum number of iterations set to 2000. 
\subsection*{Results}
\begin{figure*}[t!]
    \centering
    \includegraphics[width=1\linewidth]{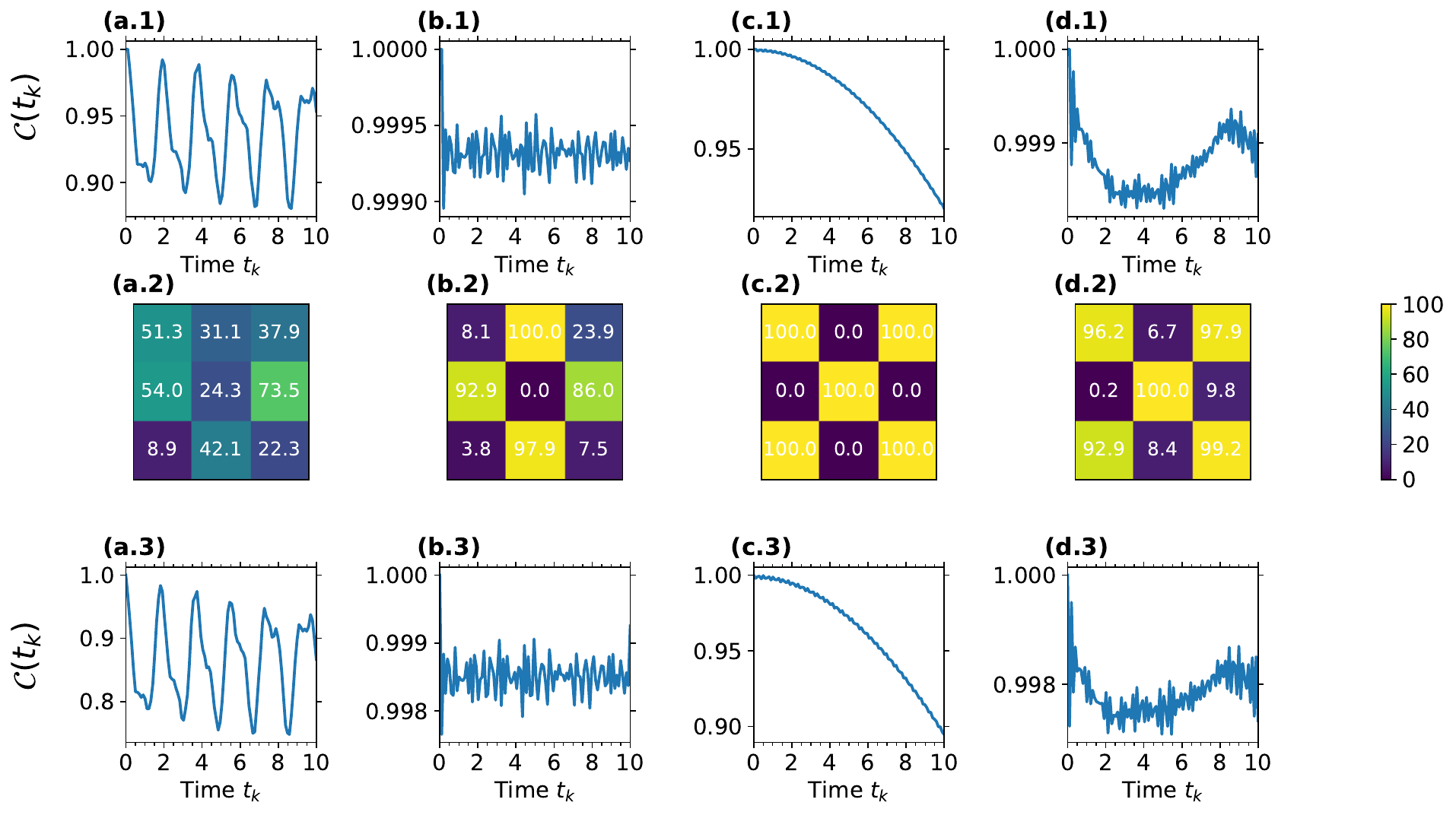}
    \caption{Time evolution of the autocorrelation function (Eq.~\eqref{eq:autocorrelationfunction}) for different sets of on-site energies $W_i$ on a $3 \times 3$ lattice. Panels (a) correspond to random on-site parameters, (b) to optimized lattice, (c) to the perfect chessboard lattice, and (d) to the perturbed chessboard lattice (see text).  For each configuration, the upper panels (x.1) show the time dependant autocorrelation function for the single-particle case, the middle panels (x.2) display the corresponding lattice structure, and the lower panels (x.3) present the time dependant autocorrelation function for the FHS. The optimization procedure is performed at the single-particle level by minimizing the cost function defined in Eq.~\eqref{eq:costfunction}.
}
    \label{fig:singleparticle}
\end{figure*}

\begin{figure}[t!]
    \centering
    \includegraphics[width=0.7\linewidth]{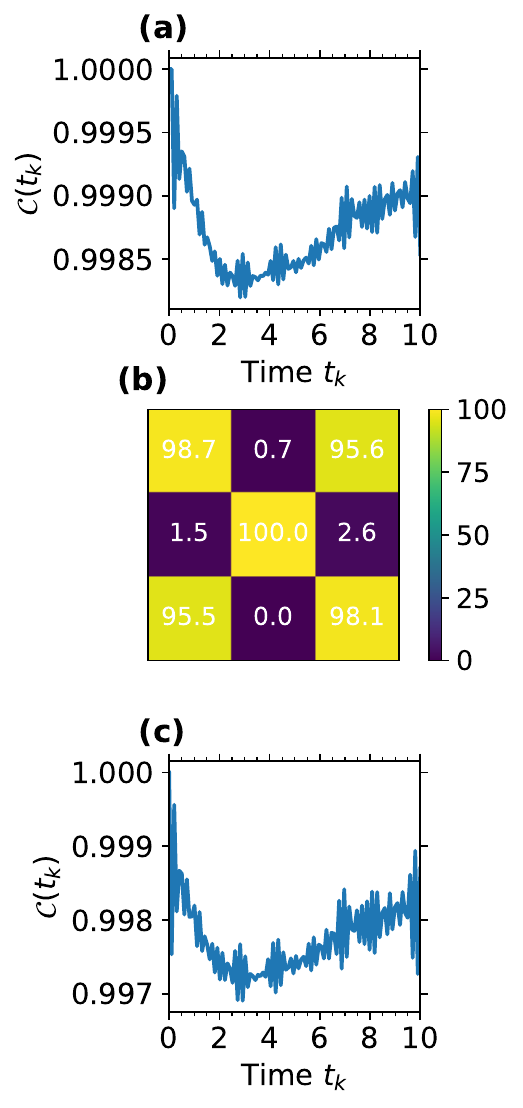}
    \caption{Time evolution of the autocorrelation function (Eq.~\eqref{eq:autocorrelationfunction}) for different sets of on-site energies $W_i$ on a $3 \times 3$ lattice. Panel (a) shows the time-dependent autocorrelation function in the full Hilbert space (FHS), panel (b) displays the corresponding lattice, and panel (c) presents the autocorrelation function in the single-particle case. The optimization is performed at the FHS level by minimizing the cost function defined in Eq.~\eqref{eq:costfunction2}.
}
\label{fig:FHS}
\end{figure}

We show various plots of the site-averaged autocorrelation function $\mathcal{C}(t)$ for different on-site energy configurations  in Fig.~\ref{fig:singleparticle}. Panel (a.1) corresponds to a randomly generated realization of $\{W_i\}$. Panels (b.1) corresponds to the optimized set $\{W_i\}$. The optimized parameters were found by minimising the cost function  Eq.\eqref{eq:costfunction}. Panel (c.1) corresponds to a chessboard pattern with staggered high and low on-site energy values. Panel (d.1) corresponds to a chessboard pattern with a small random component added to it.

In all cases, $\mathcal{C}(t)$ starts at one when $t=0$ and decreases when we start increasing $t$. The different on-site energy configurations lead to different dynamic patterns in the temporal evolution of $\mathcal{C}(t)$. Importantly, the optimized configuration leads to a rapid stabilization with only small fluctuations around a higher value ($\sim 0.99925$) of $\mathcal{C}(t)$ compared to the other cases.

This result indicates much stronger localization. Panel (b.2) shows that the optimized configuration generally resembles the chessboard pattern shown in Panel (c.2), with an irregular small component added to the chessboard pattern. To elucidate the structure of the optimized configurations, we further show the correlation functions obtained for a perfect chessboard pattern in Panel (c.1) and for a perturbed chessboard, as shown in  Panel (d.1).  
The perturbed chessboard is obtained by taking the perfect chessboard pattern and adding to it a disordered on-site energy component: for even sites, $W_i$ is randomly selected in the range $[90,100]$, whereas for odd sites, $W_i$ is randomly selected in the range $[0,10]$.  In addition, we enforce a minimal energy spacing between all pairs of sites, i.e.
\[
|W_i - W_j| > 0.1 \quad \text{for all } i \neq j,
\]
in order to avoid accidental near-degeneracies in the spectrum.

While the chessboard pattern maximizes nearest-neighbor detunings, it does not suppress transport efficiently, as evidenced by the slow but accelerating decay of $C(t)$. This steady decay indicates that suppressing first-order processes alone is not sufficient to achieve strong localization, since second-order effective hopping processes are not suppressed at all with the perfect chessboard pattern. 
Interestingly, introducing small perturbations to the chessboard pattern restores strong localization, keeping $C(t)$ close to unity over time. In fact, $\mathcal{C}(t)$ remains at the 0.999 level in this case, not much below the level obtained with the optimized energy configuration. The perturbation breaks residual resonances associated with the higher-order hopping processes, thereby reducing the effective connectivity of the lattice beyond nearest neighbors. We will derive this result using a more systematic perturbation-theory approach in Section.\ref{sec:pertubation}.

We now turn to the full qubit Hilbert space and investigate whether the localization behavior for the different configurations resembles that obtained for the single-particle problem.  As shown in Panel (b.3), the parameters optimized for the single particle space lead to a strong suppression of the dynamics in the full Hilbert space of the qubit lattice, indicating that the system remains close to a diagonal evolution in the computational basis and we observe the same behaviour for the perturbed chessboard. Importantly, the comparison between the single-particle and FHS optimizations suggests that similar interference mechanisms are at play in both cases. In particular, we observe that configurations optimized in the single-particle sector retain a strong suppression of transport when used in the full Hilbert space of a qubit lattice. This consistency suggests that the qubit lattice in some ways behaves more like a system of non-interacting particles rather than a system of interacting particles: the many-body dynamics effectively decomposes into independent single-particle contributions, so that localization optimized at the single-particle level is preserved in the full Hilbert space.

To further assess the consistency between the single-particle and FHS descriptions, we perform an independent optimization of the parameters $\{W_i\}$ directly in the full Hilbert space of the qubit lattice, using the cost function defined in Eq.~\eqref{eq:costfunction2}, which is more natural for the FHS description and does not rely on any single-particle reduction. As shown in Fig.~\ref{fig:FHS}, the optimized parameters again lead to strong localization, both in the full Hilbert space panel [(a.1)] and in the single-particle sector [panel (a.3)]. This result confirms that the localization mechanism is similar in the two problems, in spite of the huge difference in their Hilbert space sizes. 

While the dynamics in the full qubit lattice Hilbert space is exponentially more complex, it remains constrained by these same effective coupling structures. This suggests that optimization at the single-particle level can already capture key ingredients relevant for many-body localization, although a full quantitative correspondence requires the additional analysis discussed above.
These results point towards a scalable design strategy for larger systems, where optimizing local parameters in a reduced setting can provide a useful starting point for controlling transport in the full many-body dynamics.

\section*{Perturbation Theory Analysis}
\label{sec:pertubation}
Now, we will use perturbation theory to explain why the perturbed chessboard is the pattern which maximize the localization in this square lattice.

The tight binding  Hamiltonian can be decomposed as a diagonal part $H_0$ and an off-diagonal
hopping term $V$,
\begin{eqnarray}
    \hat{H}_0 &=& \sum_i W_i \hat{a}_i^\dagger \hat{a}_i,\\
    \hat{V} &=& \sum_{\langle i,j\rangle} J_{ij}
    \left(\hat{a}_j^\dagger \hat{a}_i + \mathrm{h.c.}\right).
\end{eqnarray}
In the single-particle subspace, the eigenstates of $H_0$ are the single-site localized basis states $\{|i\rangle\}$ with energies $E_i = W_i$. In this section, we treat $V$ as a perturbation that couples these localized states.

\paragraph*{First-order processes.}
To first order in $V$, the transition amplitude between two neighboring
sites $i$ and $j$ is proportional to the hopping matrix element
$J_{ij}$, but its efficiency is controlled by the detuning
\begin{equation}
    \Delta_{ij} = W_i - W_j.
\end{equation}
In the interaction picture with respect to $H_0$, the matrix element
acquires a phase factor $e^{\mathrm{i}\Delta_{ij} t}$, so that transitions
between $|i\rangle$ and $|j\rangle$ are suppressed when $|\Delta_{ij}|$ is large
(off-resonant coupling).

More precisely, time-independent perturbation theory shows that the transition
amplitude is controlled by the dimensionless ratio
\begin{equation}
    \epsilon_{ij} = \frac{J_{ij}}{\Delta_{ij}},
\end{equation}
which becomes small in the large-detuning regime $|\Delta_{ij}| \gg |J_{ij}|$.
As a result, the transition probability scales as $|\epsilon_{ij}|^2$, and
off-resonant couplings are strongly suppressed. 
Furthermore, as the results of the previous section show, the ratios $J_{ij}/\Delta_{ij}$ govern qubit decoupling as well, in spite of its vastly different Hilbert space size.

\paragraph*{Second-order processes.}
Even if all nearest-neighbor transitions are strongly detuned, a particle
can still become delocalized via virtual second-order processes $i\to k \to j$
through an intermediate site $k$. Standard second-order perturbation
theory yields an effective coupling
\begin{equation}
    J_{i\to j}^{(2,\mathrm{eff})} =
    \sum_k \frac{\langle j|\hat{V}|k\rangle \langle k|\hat{V}|i\rangle}
    {W_i - W_k},
    \label{eq:Jeff_2nd}
\end{equation}
In our lattice, $\langle
j|\hat{V}|k\rangle = J_{jk}$ and $\langle k|\hat{V}|i\rangle = J_{ki}$,
so that each virtual path $\{i \to k \to j\}$ contributes an amplitude of
order
\begin{equation}
    J_{i\to j}^{(2,\mathrm{eff})} \sim
    \frac{J_{jk} J_{ki}}{W_i - W_k}
    = \frac{J_{jk} J_{ki}}{\Delta_{ik}}.
\end{equation}
The corresponding transition probability scales as
$|J_{i\to j}^{(2,\mathrm{eff})}|^2 \propto 1/|\Delta_{ik}|^2$. Thus,
even when nearest neighbors are detuned from each other, transport to next-nearest neighbors can proceed if the associated detunings are small.

\begin{figure}
    \centering
    \includegraphics[width=0.9\linewidth]{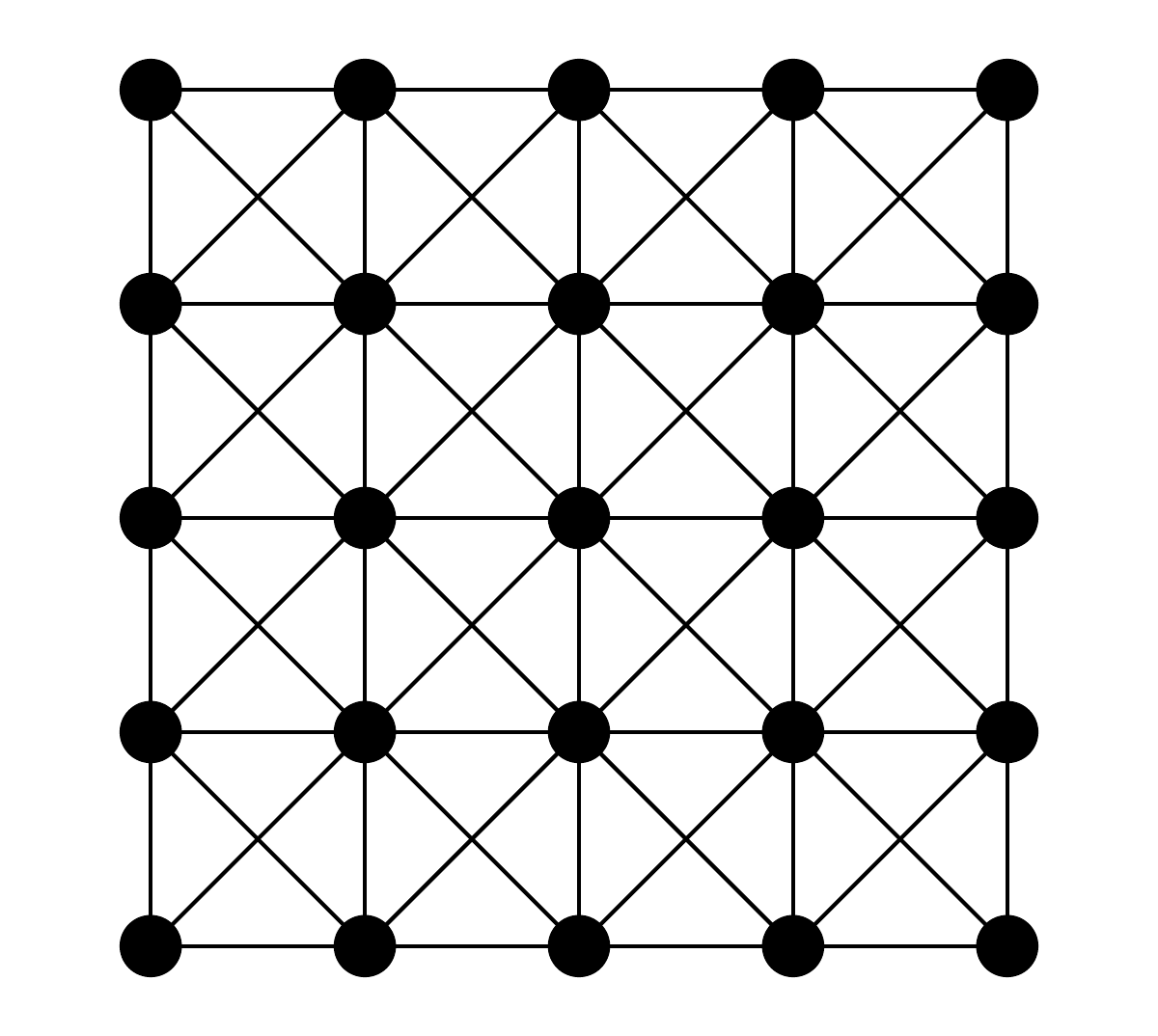}
    \caption{
Illustration of the effective connectivity of a $5 \times 5$ lattice to second order in perturbation theory. In addition to nearest-neighbor couplings, second-order processes generate effective next-nearest-neighbor connections, allowing nodes without a direct link to be connected via two-step hopping paths.  For clarity, the graph represents the effective connectivity induced by these higher-order processes and does not correspond to the bare geometry of the underlying square lattice (see text).}
    \label{fig:connectivity}
\end{figure}

\paragraph*{Graph connectivity and detunings.}
These perturbative expressions suggest a natural notion of
\emph{connectivity} for the lattice: each nearest-neighbor edge $(i,j)$
carries an effective weight proportional to $1/|\Delta_{ij}|$, and each
second-order path $(i\to k\to j)$ carries a weight proportional to
$1/|\Delta_{ik}|$.  We illustrate these with a scheme of connectivity of this lattice in Fig.\ref{fig:connectivity}. 

To suppress both first- and second-order transport,
we therefore seek configurations $\{W_i\}$ that maximize the relevant
detunings,
\begin{eqnarray}
    |\Delta_{ij}| = |W_i - W_j| \quad \text{(nearest neighbors)},\qquad \\
    |\Delta_{ik}| = |W_i - W_k| \quad \text{(next-nearest neighbors)}.
\end{eqnarray}

Instead of working directly with the inverse weights, we employ a
cost function that rewards large detunings by penalizing their inverse
square.  These perturbative processes naturally define an effective connectivity between lattice sites. We introduce an effective weighted graph where each edge ((i,j)) carries a weight
\begin{equation}
w_{ij}
\sim
\left|
\frac{J_{ij}}{\Delta_{ij}}
\right|^2
+
\sum_k
\left|
\frac{J_{ik}J_{kj}}
{\Delta_{ik}\Delta_{kj}}
\right|^2 .
\end{equation}
Localization to second order in perturbation theory can then be interpreted as the suppression of transport on this effective graph. In this picture, strong localization corresponds to minimizing the total connectivity of the graph.

Motivated by this perturbative structure, we define a cost function that penalizes effective connectivity:
\begin{eqnarray}
L_{\mathrm{pert}}
=
\sum_{i=1}^{n}
\sum_{j\in \mathrm{NN}(i)}
\left(
\frac{1}{|W_i-W_j|}
\right)^2
+ \nonumber \\ 
\alpha
\sum_{i=1}^{n}
\sum_{j\in \mathrm{NNN}(i)}
\left(
\frac{1}{|W_i-W_j|}
\right)^2 ,
\label{eq:Lpert_general}
\end{eqnarray}

where the first sum runs over nearest-neighbor pairs and the second over
next-nearest neighbors on the lattice. The parameter $\alpha$ sets the
relative importance of second-order paths compared to direct hopping.
We choose $\alpha = 0.01$, corresponding to a typical
ratio of $|J|/|\Delta|$ for the parameters considered. Minimizing
$L_{\text{pert}}$ therefore amounts to maximizing a weighted sum of
nearest- and next-nearest detunings, which directly reduces the
effective first- and second-order connectivities encoded in
Eq.~\eqref{eq:Jeff_2nd}. 

\begin{figure*}[t!]
    \centering
    \includegraphics[width=0.5\linewidth]{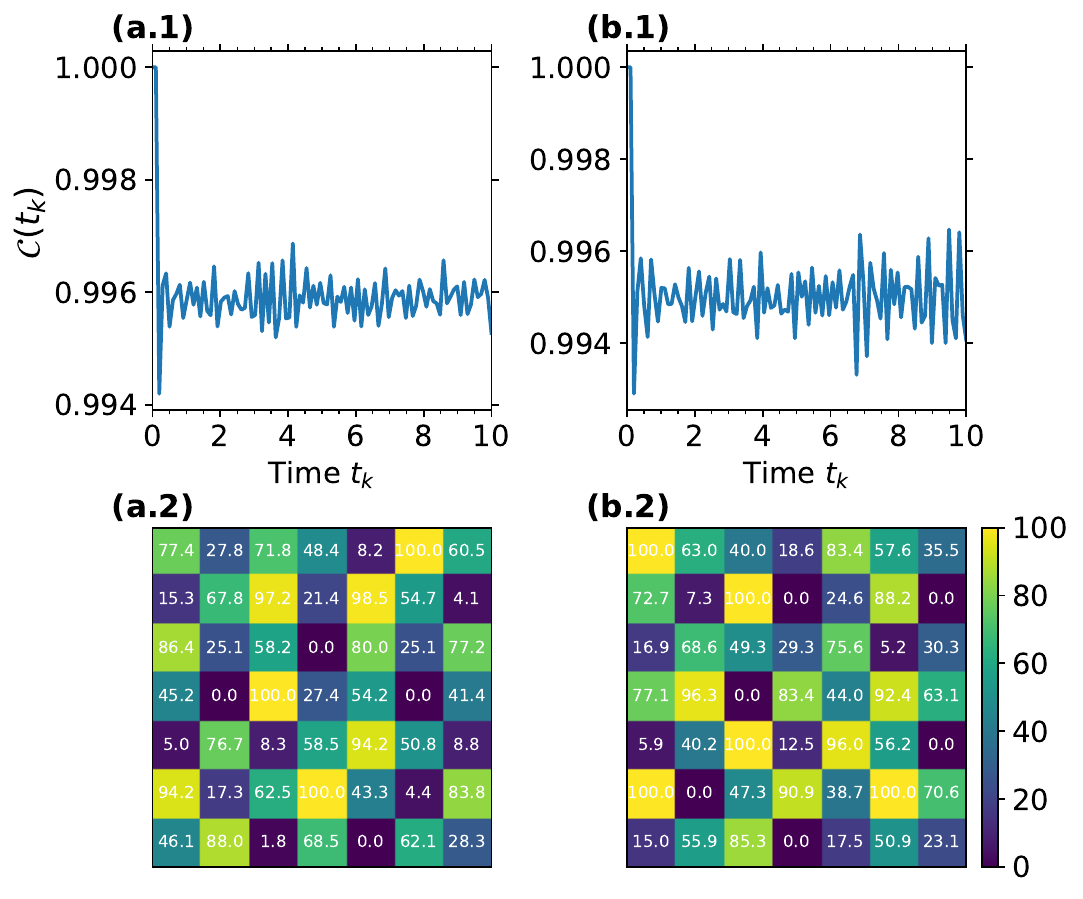}
    \caption{Time evolution of the autocorrelation function (Eq.~\eqref{eq:autocorrelationfunction}) for a $7 \times 7$ lattice with on-site energies $W_i$ in the single-particle case. Panel (a) shows the time-dependent autocorrelation function for the optimized lattice obtained by minimizing the cost function defined in Eq.~\eqref{eq:costfunction}, while panel (a.2) displays the corresponding lattice configuration.  Panel (b) presents the autocorrelation function obtained from the perturbative optimization based on Eq.~\eqref{eq:Lpert_general}, and panel (b.2) shows the associated lattice configuration.}
    \label{fig:perturbationtheory}
\end{figure*}

Crucially, $L_{\text{pert}}$ captures the qualitative dependence on the detunings: configurations with large nearest-neighbor detunings, together with additional second-order detuning, yield small values of $L_{\text{pert}}$ and
are numerically observed to exhibit strong localization, both in the single-particle sector and in the full Hilbert space. This perturbative cost function therefore provides a fast surrogate that can be used to explore larger lattices, with many particles, as illustrated in Fig.~\ref{fig:perturbationtheory}. In Fig.~\ref{fig:perturbationtheory}(a), the parameters obtained by minimizing $L_{\text{pert}}$ lead to a  stronger localization than a typical
configuration with random on-site energies $W_i$, shown in Fig.~\ref{fig:perturbationtheory}(b).

\section*{Conclusion}

We have shown that localization in two-dimensional lattices governed by
the tight binding  Hamiltonian can be significantly enhanced by 
systematically optimizing the on-site energies $W_i$.
Using a dynamical localization parameter and direct numerical optimization,
we identified spatial patterns of nearest- and next-nearest-neighbor detunings that strongly suppress both
first-and second-order hopping processes.  
In particular, simple regular structures, such as chessboard
patterns, are not optimal.  
Instead, small tailored perturbations or fully optimized configurations
produce much stronger localization, as confirmed by the correlation
function $C(t)$.

Using perturbation theory, we derived an improved cost function,
$L_{\text{pert}}$, which captures the effective connectivity of the
lattice by incorporating both nearest- and next-nearest-neighbor
detunings.  
Minimizing this cost function naturally leads to on-site energies that
significantly suppress the effective couplings $J^{\mathrm{eff}}_{i\to j}$ up to second
order, which explains why the optimized patterns achieve superior
localization. We further demonstrated that these optimized parameters remain efficient
when transferred to the full $2^n$-dimensional Hilbert space of $n$
qubits, indicating that single-particle optimization already contains
the relevant physical constraints for multi-qubit systems.  
This result suggests a practical way to engineer localized dynamics and
suppress undesired transitions, like error propagation, in quantum processors.

Our approach provides a practically simple and physically intuitive method to design on-site energies in engineered quantum systems to maximize localization. The framework can be extended to larger and more complex lattices, higher-order processes, applications such as crosstalk suppression, quantum simulation, and robust qubit architectures.

\section*{Acknowledgement}
 We would like to thank Andreas Buchleitner and Pedram Roushan for useful discussions. M.B. thanks the Georg H.~Endress Stiftung for funding and support. S.A. acknowledges support from Japan's Ministry of Education, Culture, Sports, Science and Technology (MEXT) Quantum Leap Flagship Program Grant Number JPMXS0120319794.
\bibliographystyle{apsrev4-2}
\bibliography{biblio}

\end{document}